\documentstyle[12pt,graphicx,pstricks,braket,amsmath,amssymb,enumitem,
                   float,cite,tabu,hyperref,epstopdf,subfigure, float,tikz]{article}
                    \usetikzlibrary{arrows,matrix,positioning}
\numberwithin{equation}{section}

\newcommand{\hi}[1]{}

\begin{document}

\def\AEF{A.E. Faraggi}

\def\JHEP#1#2#3{{JHEP} {\textbf #1}, (#2) #3}
\def\vol#1#2#3{{\bf {#1}} ({#2}) {#3}}
\def\NPB#1#2#3{{\it Nucl.\ Phys.}\/ {\bf B#1} (#2) #3}
\def\PLB#1#2#3{{\it Phys.\ Lett.}\/ {\bf B#1} (#2) #3}
\def\PRD#1#2#3{{\it Phys.\ Rev.}\/ {\bf D#1} (#2) #3}
\def\PRL#1#2#3{{\it Phys.\ Rev.\ Lett.}\/ {\bf #1} (#2) #3}
\def\PRT#1#2#3{{\it Phys.\ Rep.}\/ {\bf#1} (#2) #3}
\def\MODA#1#2#3{{\it Mod.\ Phys.\ Lett.}\/ {\bf A#1} (#2) #3}
\def\RMP#1#2#3{{\it Rev.\ Mod.\ Phys.}\/ {\bf #1} (#2) #3}
\def\IJMP#1#2#3{{\it Int.\ J.\ Mod.\ Phys.}\/ {\bf A#1} (#2) #3}
\def\nuvc#1#2#3{{\it Nuovo Cimento}\/ {\bf #1A} (#2) #3}
\def\RPP#1#2#3{{\it Rept.\ Prog.\ Phys.}\/ {\bf #1} (#2) #3}
\def\APJ#1#2#3{{\it Astrophys.\ J.}\/ {\bf #1} (#2) #3}
\def\APP#1#2#3{{\it Astropart.\ Phys.}\/ {\bf #1} (#2) #3}
\def\EJP#1#2#3{{\it Eur.\ Phys.\ Jour.}\/ {\bf C#1} (#2) #3}
\def\etal{{\it et al\/}}
\def\notE6{{$SO(10)\times U(1)_{\zeta}\not\subset E_6$}}
\def\E6{{$SO(10)\times U(1)_{\zeta}\subset E_6$}}
\def\highgg{{$SU(3)_C\times SU(2)_L \times SU(2)_R \times U(1)_C \times U(1)_{\zeta}$}}
\def\highSO10{{$SU(3)_C\times SU(2)_L \times SU(2)_R \times U(1)_C$}}
\def\lowgg{{$SU(3)_C\times SU(2)_L \times U(1)_Y \times U(1)_{Z^\prime}$}}
\def\SMgg{{$SU(3)_C\times SU(2)_L \times U(1)_Y$}}
\def\Uzprime{{$U(1)_{Z^\prime}$}}
\def\Uzeta{{$U(1)_{\zeta}$}}

\newcommand{\cc}[2]{c{#1\atopwithdelims[]#2}}
\newcommand{\bev}{\begin{verbatim}}
\newcommand{\beq}{\begin{equation}}
\newcommand{\ba}{\begin{eqnarray}}
\newcommand{\ea}{\end{eqnarray}}

\newcommand{\beqa}{\begin{eqnarray}}
\newcommand{\beqn}{\begin{eqnarray}}
\newcommand{\eeqn}{\end{eqnarray}}
\newcommand{\eeqa}{\end{eqnarray}}
\newcommand{\eeq}{\end{equation}}
\newcommand{\beqt}{\begin{equation*}}
\newcommand{\eeqt}{\end{equation*}}
\newcommand{\Eev}{\end{verbatim}}
\newcommand{\bec}{\begin{center}}
\newcommand{\eec}{\end{center}}
\newcommand{\bes}{\begin{split}}
\newcommand{\ees}{\end{split}}
\def\ie{{\it i.e.~}}
\def\eg{{\it e.g.~}}
\def\half{{\textstyle{1\over 2}}}
\def\nicefrac#1#2{\hbox{${#1\over #2}$}}
\def\third{{\textstyle {1\over3}}}
\def\quarter{{\textstyle {1\over4}}}
\def\m{{\tt -}}
\def\mass{M_{l^+ l^-}}
\def\p{{\tt +}}

\def\slash#1{#1\hskip-6pt/\hskip6pt}
\def\slk{\slash{k}}
\def\GeV{\,{\rm GeV}}
\def\TeV{\,{\rm TeV}}
\def\y{\,{\rm y}}

\def\l{\langle}
\def\r{\rangle}
\def\LRS{LRS  }

\begin{titlepage}
\samepage{
\setcounter{page}{1}
\rightline{LTH--1093}
\vspace{1.5cm}

\begin{center}
 {\Large \bf 
The Observed Diphoton Excess in F-theory Inspired Heterotic String-Derived Model}
\end{center}

\begin{center}

{\large
Johar M. Ashfaque$^\spadesuit$\footnote{email address: jauhar@liv.ac.uk}
}\\
\vspace{1cm}
$^\spadesuit${\it  Dept.\ of Mathematical Sciences,
             University of Liverpool,
         Liverpool L69 7ZL, UK\\}
\end{center}

\begin{abstract}
The production and the subsequent decay of the SM singlet via heavy vector--like colour triplets and electroweak doublets in one--loop diagrams can shed light on the recent observation of diphoton excess at the LHC. In this paper, the $E_6$ GUT is considered in the F-theory setting where the $E_6$ is broken by making use of the spectral cover construction and by turning on the hypercharge gauge flux. This paper is based on the results presented in \cite{Athanasopoulos:2014bba, Faraggi:2016xnm, Ashfaque:2016jha} which will be reviewed briefly. Here, by following the F-theory approach, akin to \cite{Karozas:2016hcp, Leontaris:2016wsy, Das:2016xuc}, we present a study of the flipped $SO(10)$ model embedded completely in the $E_{6}$ GUT but with a different accommodation of the SM representations in the ${\bf{27}}$ of $E_{6}$.
\end{abstract}
\smallskip}
\end{titlepage}

\section{Introduction}
The recent observation of diphoton excess at the LHC reported by the ATLAS \cite{atlas} and CMS \cite{cms} collaborations can be understood through the production and the subsequent decay of a SM singlet via heavy vector--like colour triplets and electroweak doublets in one--loop diagrams has sparked significant interest \cite{flurry}. 

The Type IIB superstring theory admits a class of non-perturbative compactifications that go by the name of F-theory \cite{Vafa:1996xn, Morrison:1996na, Morrison:1996pp}. To break a GUT symmetry in F-theory models, one can either use Wilson lines \cite{Beasley:2008k,Chung:2009ib} or introduce a supersymmetric $U(1)$ flux corresponding to a fractional line bundle \cite{Beasley:2008kw, Marsano:2009ym, Marsano:2009gv, Marsano:2009wr, Chen:2010tp, Chen:2010ts}. In local models, an Abelian or a non-Abelian gauge flux of the rank higher than two may be turned on on the bulk to break the gauge group \cite{Beasley:2008kw}. There are two kinds of rank three fluxes, $U(1)^3$ and $SU(2)\times U(1)^2$, both embedded in the $E_6$ gauge group with commutants including the Standard Model (SM) gauge structure. For simplicity, we will focus on $U(1)^3$. 

The aim of this paper is to present a study of the flipped $SO(10)$ model embedded completely in the $E_{6}$ GUT but with a different accommodation of the SM representations in the ${\bf{27}}$ of $E_{6}$ in a string-derived heterotic low-energy effective model constructed in the free fermionic formulation. The chiral spectrum of the model will be seen to form complete $E_6$ representations.

\section{A String-Derived Low-Energy Effective Model} 
The string-derived model in \cite{frzprime} was 
constructed in the free fermionic formulation \cite{fff} of the four-dimensional heterotic string. The complete details along with the the massless spectrum  and the superpotential can be found in \cite{frzprime} and are therefore omitted here. 
The chiral spectrum of the model, \cite{frzprime}, forms complete $E_6$ representations, whereas the additional vector--like multiplets may reside in incomplete multiplets. This is in fact an additional important property of the model,  which affects compatibility with the gauge coupling data. Space-time vector bosons are obtained solely from the untwisted sector and generate the observable and hidden gauge symmetries:
\beqn
{\rm observable} ~: &~~~~~~~~SO(6)\times SO(4) \times 
U(1)_1 \times U(1)_2\times U(1)_3 \nonumber\\
{\rm hidden}     ~: &SO(4)^2\times SO(8)~.~~~~~~~~~~~~~~~~~~~~~~~\nonumber
\eeqn
The $E_6$ combination being
\beq
U(1)_\zeta = U(1)_1+U(1)_2+U(1)_3 ~,
\label{u1zeta}
\eeq
which is anomaly free whereas the orthogonal combinations of $U(1)_{1,2,3}$
are anomalous. 
The model also contains vector--like states that transform
under the hidden $SU(2)^4\times SO(8)$ group factors, with charges 
$Q_\zeta=\pm1$ or $Q_\zeta=0$. 

Here we  consider the PS breaking scale to be in the vicinity of the string scale where the VEVs of the heavy Higgs fields that break the PS gauge group leave an unbroken $U(1)_{Z^\prime}$ symmetry given by
\beq
U(1)_{{Z}^\prime} ~=~
{1\over {2}} U(1)_{B-L} -{2\over3} U(1)_{T_{3_R}} - {5\over3}U(1)_\zeta
~\notin~ SO(10),
\label{uzpwuzeta}
\eeq
which can be found to remain unbroken down to low scales provided that $U(1)_\zeta$ is
anomaly free. 
Cancellation of the anomalies requires
that the additional vector--like quarks and leptons, 
that arise from the ${\bf{10}}$ of $SO(10)$, 
as well as the $SO(10)$ singlet in the ${\bf{27}}$ of $E_6$, remain in the light spectrum. The spectrum below the PS breaking scale is displayed schematically 
in table \ref{table27rot}. 
The spectrum is taken to be supersymmetric down to the TeV scale.
As in the MSSM,
compatibility of gauge coupling unification with the
experimental data requires the existence of one vector--like 
pair of Higgs doublets, beyond the number of vector--like triplets. 
\begin{table}[H]
\noindent 
{\small
\begin{center}
{\tabulinesep=1.2mm
\begin{tabu}{|l|cc|c|c|c|}
\hline
Field &$\hphantom{\times}SU(3)_C$&$\times SU(2)_L $
&${U(1)}_{Y}$&${U(1)}_{Z^\prime}$  \\
\hline
$Q_L^i$&    $3$       &  $2$ &  $+\frac{1}{6}$   & $-\frac{2}{3}$   ~~  \\
$u_L^i$&    ${\bar3}$ &  $1$ &  $-\frac{2}{3}$   & $-\frac{2}{3}$   ~~  \\
$d_L^i$&    ${\bar3}$ &  $1$ &  $+\frac{1}{3}$   & $-\frac{4}{3}$  ~~  \\
$e_L^i$&    $1$       &  $1$ &  $+1          $   & $-\frac{2}{3}$  ~~  \\
$L_L^i$&    $1$       &  $2$ &  $-\frac{1}{2}$   & $-\frac{4}{3}$  ~~  \\
%
\hline
$D^i$       & $3$     & $1$ & $-\frac{1}{3}$     & $+\frac{4}{3}$  ~~    \\
${\bar D}^i$& ${\bar3}$ & $1$ &  $+\frac{1}{3}$  &   ~~$~2$  ~~    \\
$H^i$       & $1$       & $2$ &  $-\frac{1}{2}$   &  ~~$~2$ ~~    \\
${\bar H}^i$& $1$       & $2$ &  $+\frac{1}{2}$   &   $+\frac{4}{3}$   ~~  \\
\hline
$S^i$       & $1$       & $1$ &  ~~$0$  &  $-\frac{10}{3}$       ~~   \\
\hline
$h$         & $1$       & $2$ &  $-\frac{1}{2}$  &  $-\frac{4}{3}$  ~~    \\
${\bar h}$  & $1$       & $2$ &  $+\frac{1}{2}$  &  $+\frac{4}{3}$  ~~    \\
\hline
$\phi$       & $1$       & $1$ &  ~~$0$         & $-\frac{5}{3}$     ~~   \\
$\bar\phi$       & $1$       & $1$ &  ~~$0$     & $+\frac{5}{3}$     ~~   \\
\hline
%
$\zeta^i$       & $1$       & $1$ &  ~~$0$  &  ~~$0$       ~~   \\
\hline
\end{tabu}}
\end{center}
}
\caption{\label{table27rot}
\it
Spectrum and
$SU(3)_C\times SU(2)_L\times U(1)_{Y}\times U(1)_{{Z}^\prime}$ 
quantum numbers, with $i=1,2,3$ for the three light 
generations. The charges are displayed in the 
normalisation used in free fermionic 
heterotic--string models. }
\end{table}

\section{The $E_{6}$ Singularity}
$$E_{8}\supset E_{6}\times SU(3)_{\perp}\rightarrow E_{6}\times U(1)^{2}_{\perp}$$
with 
$${\bf{248}} \rightarrow ({\bf{78}},{\bf{1}})+({\bf{1}},{\bf{8}})+({\bf{27}},{\bf{3}})+(\overline{{\bf{27}}},\overline{{\bf{3}}})$$

In accordance with the standard terminology, the $SU(3)_{\perp}$ factor is considered as the group `perpendicular' to the $E_{6}$ GUT divisor.
In what follows assume semi-local approach where the $E_{6}$ representations transform non-trivially under the $SU(3)_{\perp}$. In the spectral cover approach the $E_{6}$ representations are distinguished by the weights $t_{1,2,3}$ of the $SU(3)_{\perp}$ Cartan subalgebra subject to the traceless condition 
$$\sum_{i=1}^{3}t_{i}=0$$
while the $SU(3)_{\perp}$ adjoint decomposes into singlets $1_{t_{i}-t_{j}}\equiv \theta_{ij}$\footnote{We introduce the notation $(1, 8)\rightarrow \theta_{ij}$.}.

The $E_6$ content consists of three {\bf{27}}s (and $\overline{\bf{27}}$s) plus eight singlet matter curves. In terms of the
weight vectors $t_i$, $i=1,2,3$ of $SU(3)_{\perp}$ the equations of these curves are
\begin{eqnarray*}
\sum\,\,_{27}&:& t_{i}=0,\\
\sum\,\,_{1}&:& \pm(t_{i}-t_{j})=0\qquad i\neq j\,\,.\\
\end{eqnarray*}

Under the decomposition $E_{6}\rightarrow SO(10)\times U(1)_{\zeta}$, following from table \ref{table27rot}, the relevant $E_{6}$ representations decompose as follows
\beqn
{\bf{27}} & \rightarrow & {\bf{16}}_{+{1/2}} + {\bf{10}}_{-1} + {\bf{1}}_{+2},\nonumber\\
{\bf{\overline{27}}} & \rightarrow & {\bf{\overline{16}}}_{-{1/2}} + {\bf{\overline{10}}}_{+1} + {\bf{1}}_{-2}.\nonumber
\eeqn

The $E_{6}$ GUT symmetry can be broken following \cite{Hewett:1988xc} as
\begin{eqnarray*}
E_{6}&\rightarrow& SO(10)\times U(1)_{\zeta}\\&\rightarrow& [SU(5)\times U(1)_{\zeta'}]\times U(1)_{\zeta}\\ &\rightarrow& [SU(3)\times SU(2)\times U(1)_{\zeta''}]\times U(1)_{\zeta'} \times U(1)_{\zeta}
\end{eqnarray*}
and the SM representations are accommodated in the ${\bf{27}}$ of $E_{6}$ as 
$${\bf{27}} = \begin{cases} {\bf{16}}_{+\frac{1}{2}} & \mathcal{F}_{L}+\mathcal{F}_{R} = ($$Q$$,\,$$u^{c}$$, d^{c}, L, e^{c}, N)\\& \rightarrow \binom{Q\,\,\,\,u^{c}}{e^{c}}+\binom{d^{c}}{L}+N\\
{\bf{10}}_{-1}& \mathcal{D} + \mathcal{H} \\ \,\,\,{\bf{1}}_{+2} & {\mathcal{S}\rightarrow }\,\,S\end{cases}.$$ 

\section{{The Observed Diphoton Excess}}
Implementing the $\mathbb{Z}$$_{2}$ monodromy via the binomial-monomial factorization, we have 

\[\left[ 
\begin{array}{c@{}c@{}c}
 \left[\begin{array}{cc}
        \textcolor{red}{ t_{1}} & \\
          & \textcolor{red}{ t_{2}} \\
  \end{array}\right] &   \\
 & \left[\begin{array}{ccc}
                      \textcolor{red}{ t_{3}}\\
                      \end{array}\right] \\
\end{array}\right]
\]    
and the following relations between the eight singlets
\begin{eqnarray*}
\theta_{12}&=&\theta_{21}\equiv\theta_{0}\\
\theta_{23}&=&\theta_{13}\\
\theta_{32}&=&\theta_{31} \leftrightarrow S\\
\end{eqnarray*}
where we identify 750 GeV resonance $S$ with the singlet $\theta_{32}=\theta_{31}$ which couples to vector pairs.

As mentioned in \cite{Ashfaque:2016jha, frzprime}, in the low-energy regime, the superpotential provides different 
interaction terms of the singlet fields $S_i$ and $\zeta_i$ which can be extracted 
from table ~\ref{table27rot}, among them we have the following
\beqn
\label{superpot}
\lambda^{ijk}_D S_i D_j \bar D_k + 
\lambda^{ijk}_H S_i H_j \bar H_k + 
\lambda^{ij}_h S_i H_j \bar h + 
\eta^i_{ \mathcal D} \zeta_i  {\mathcal D} \bar{ \mathcal D} +  
\eta^i_{ h} \zeta_i  h \bar{h} \,.
\eeqn

\section{Conclusions}
In this paper, using Abelian fluxes to realise the $E_{6}$ GUT symmetry breaking, we presented a study of the flipped $SO(10)$ model embedded completely in the $E_{6}$ GUT but with a different accommodation of the SM representations in the ${\bf{27}}$ of $E_{6}$. Moreover, the production and the subsequent decay of the SM singlet via heavy vector--like colour triplets and electroweak doublets in one--loop diagrams can shed light on the recent LHC-observed diphoton excess, which in effect will be able to provide pivotal evidence in understanding the fundamental origins of the SM. 
\section{Acknowledgements}
J. M. A. would like to thank the String Phenomenology 2016 organisers hosted in Ioannina and String-Math 2016 organisers hosted in Paris for their warm hospitality.

\end{document}